\documentclass[12pt]{iopart}

\usepackage{iopams,psfrag,graphicx,color}  

\def\bc{\begin{center}}
\def\ec{\end{center}}

\def\beq{\begin{equation}}
\def\eeq{\end{equation}}

\begin{document}

\title[]{Gaps and tails in graphene and graphane}

\author{B. D\'ora$^1$ and K. Ziegler$^2$}

\address{$^1$Max-Planck-Institut f\"ur Physik Komplexer Systeme,\\
N\"othnitzer Str. 38, 01187 Dresden, Germany\\
$^2$Institut f\"ur Physik, Universit\"at Augsburg}
\ead{klaus.ziegler@physik.uni-augsburg.de}

\begin{abstract}
We study the density of states in monolayer and bilayer graphene in the presence of a 
random potential that breaks sublattice symmetries. While a uniform symmetry-breaking potential 
opens a uniform gap, a random symmetry-breaking potential also creates tails in the density of states. 
The latter can close the gap again, preventing the system to become an
insulator. However, for a sufficiently large gap the tails contain localized states with nonzero density
of states. These localized states allow the system to conduct at nonzero temperature via variable-range 
hopping. This result is in agreement with recent experimental observations in graphane by Elias {\it et al.}.
\end{abstract}

\maketitle


\section{Introduction}

Graphene is a single sheet of carbon atoms that is forming a honeycomb lattice. 
A graphene monolayer as well as a stack of two graphene sheets (i.e. a graphene bilayer) are semimetals 
with remarkably good conducting properties \cite{novoselov05,zhang05,geim07}. 
These materials have been experimentally realized with external gates,
which allow for a continuous change in the charge-carrier density. There exists a non-zero minimal conductivity
at the charge neutrality point. Its value is very robust and almost unaffected by disorder or 
thermal fluctuations \cite{geim07,tan07,chen08,morozov08}.

Many potential applications of graphene require an electronic gap to switch between conducting and
insulating states. A successful step in this direction has been achieved by recent 
experiments with hydrogenated graphene (graphane) \cite{elias08} and with gated bilayer graphene 
\cite{ohta06,oostinga08,gorbachev08}. These experiments take advantage of the fact 
that the breaking of a discrete symmetry of the lattice system opens a gap in the electronic 
spectrum at the Fermi energy. In the case of monolayer graphene (MLG), a staggered potential 
that depends on the sublattice of the honeycomb lattice plays the role of such symmetry-breaking 
potential (SBP). For bilayer graphene (BLG) a gate potential that 
distinguishes between the two graphene layers plays a similar role.

With these opportunities one enters a new field in graphene, where one can switch 
between conducting and insulating regimes of a two-dimensional material, 
either by a chemical process (e.g. oxidation or hydrogenation) or by applying an 
external electric field \cite{castro08}.

The opening of a gap can be observed experimentally either by a direct measurement of the
density of states (e.g., by scanning tunneling microscopy \cite{li09}) or indirectly by measuring 
transport properties.
In the gapless case we observe a metallic conductivity
$
\sigma\propto \rho D
$,
where $D$ is the diffusion coefficient (which is proportional to the scattering time) and $\rho$ 
is the density of states (DOS). This gives typically
a conductivity of the order of $e^2/h$.
The gapped case, on the other hand, has a strongly temperature-dependent conductivity due to 
thermal activation of charge carriers \cite{mott90}
\begin{equation}
\sigma(T) = \sigma_0 e^{-T_0/T}
\end{equation}
with some characteristic temperature scale $T_0$ which depends on the underlying model.
A different behavior was found experimentally in the insulating phase of graphane \cite{elias08}:
\beq
\sigma(T)\approx\sigma_0e^{-(T_0/T)^{1/3}} \  ,
\label{vrh}
\eeq
which is known as 2D variable-range hopping \cite{mott69}. This behavior
indicates the existence of well-separated localized states, even at the charge-neutrality
point, where the parameter $T_0$ depends on the DOS at the Fermi energy
$E_F$ as $T_0\propto 1/\rho(E_F)$.

The experimental observation of a metal-insulator transition in graphane raises two questions:
(i) what are the details that describe the opening of a gap and (ii) what is the DOS
in the insulating phase? 
In this paper we will focus on the mechanism of the gap opening due to a SBP in MLG and BLG.
It is crucial for our study that the SBP is not uniform in the realistic 
two-dimensional material. One reason for the latter is the fact that 
graphene is not flat but forms ripples \cite{morozov06,meyer07,castroneto07b}. 
Another reason is the incomplete
coverage of a graphene layer with hydrogen atoms in the case of graphane
\cite{elias08}. The spatially fluctuating SBP leads to interesting effects, including
a second-order phase transition due to spontaneous breaking of a discrete symmetry
and the formation of Lifshitz tails.

\section{Model}
\label{sect2b}

Quasiparticles in MLG or in BLG are described in tight-binding 
approximation by a nearest-neighbor hopping Hamiltonian 
\beq
{\bf H}=-{\sum_{<r,r'>}}t_{r,r'} c^\dagger_r c_{r'}
+\sum_r V_r c^\dagger_r c_r +h.c.,
\label{ham00}
\eeq
where $c_r^\dagger$ ($c_r$) are fermionic creation (annihilation) operators at lattice site $r$.
The underlying lattice structure is either a honeycomb lattice (MLG) or two honeycomb lattices 
with Bernal stacking (BLG) \cite{castro08,mccann06b}. We have an intralayer hopping rate $t$ and 
an interlayer hopping rate $t_\perp$ for BLG. 
There are different forms of the potential $V_r$, depending on whether we consider MLG or BLG.
Here we begin with potentials that are uniform on each sublattice, whereas random fluctuations 
are considered in subsection \ref{randomfluct}.

\subsection{MLG}

$V_r$ is a staggered potential with $V_r=m$ on sublattice A and $V_r=-m$ on sublattice B.
This potential obviously breaks the sublattice symmetry of MLG. Such a staggered potential 
can be the result of chemical absorption of non-carbon atoms in MLG (e.g. oxygen or hydrogen 
\cite{elias08}). 
A consequence of the symmetry breaking is the formation of a gap $\Delta_g=m$:
The spectrum of MLG consists of two bands with dispersion
\beq
E_k=\pm\sqrt{m^2+\epsilon_k^2}  ,
\eeq
where
\beq
\epsilon_k^2=t^2[3+2\cos k_1+4\cos(k_1/2)\cos(\sqrt{3}k_2/2)]
\label{specmlg}
\eeq
for lattice spacing $a=1$.

\subsection{BLG}

$V_r$ is a biased gate
potential that is $V_r=m$ ($V_r=-m$) on the upper (lower) graphene sheet. 
The potential in BLG has been realized as an 
external gate voltage, applied to the two layers of BLG \cite{ohta06}.
The spectrum of BLG consists of four bands \cite{castro08} with two low-energy bands 
\beq
E_k^-(m)=\pm\sqrt{\epsilon_k^2+t_\perp^2/2+m^2-\sqrt{t_\perp^4/4+(t_\perp^2+4m^2)\epsilon_k^2}} \ ,
\eeq
where $\epsilon_k$ is the monolayer dispersion of Eq. (\ref{specmlg}), and two high-energy bands
\beq
E_k^+(m)=\pm\sqrt{\epsilon_k^2+t_\perp^2/2+m^2+\sqrt{t_\perp^4/4+(t_\perp^2+4m^2)\epsilon_k^2}} \ .
\eeq
The spectrum of the low-energy bands has nodes for $m=0$ where $E_k^-(0)$ vanishes in a $(k-K)^2$ 
manner, where $K$ is the position of the nodes, which are the same as those of a single layer. 
For small $m\ll t_\perp$, a mexican hat structure develops around $k=K$, with local extremum in the low-energy band at $E_k^-(m)=\pm m$, 
and a global minimum/maximum in the upper/lower low energy band at
$E_k^-(m)=mt_\perp/\sqrt{t_\perp^2+4m^2}$. 

\begin{figure}[h!]
\centering
\psfrag{x}[t][b][1][0]{$|k|-K$}
\psfrag{y}[b][t][1][0]{$E_k$}
\psfrag{bi}[][][1][0]{\color{red} bilayer}
\psfrag{mono}[][][1][0]{\color{blue} monolayer}
\includegraphics[width=7cm,height=7cm]{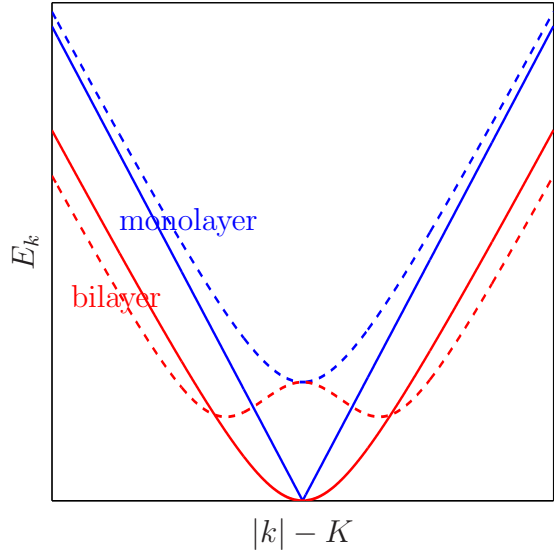}
\caption{The energy spectra of MLG (blue) and BLG (red) are shown, with and without a gap (dashed and solid line, respectively) for positive energies. 
Note the characteristic mexican hat structure of gapped 
BLG.
\label{figspecmlg}}
\end{figure}

For small gating potential $V_r=\pm m$
we can expand $E_k^-(m)$ under the square root near the nodes and get
\begin{equation}
 E_k^-(m)\sim \pm\sqrt{[1-4\epsilon_k^2t_\perp(t_\perp^2+4\epsilon_k^2)^{-1/2}]m^2+E_k^-(0)^2} \ .
 \end{equation}
$t_\perp$ apparently reduces the gap. 
Very close to the nodes we can approximate the factor in front of $m^2$ by 1 and obtain an expression similar
to the dispersion of MLG: $ E_k^-(m)\sim \pm\sqrt{m^2+E_k^-(0)^2}$. Here we notice the absence of the mexican hat structure in this approximation.
The resulting spectra for MLG and BLG are shown in Fig. \ref{figspecmlg}.

\subsection{Low-energy approximation}

The two bands in MLG and the two low-energy bands in BLG represent a spinor-1/2 wave
function. This allows us to expand the corresponding Hamiltonian in terms of
Pauli matrices $\sigma_j$ as
\beq
H=h_1\sigma_1+h_2\sigma_2+m\sigma_3 \ .
\label{ham01}
\eeq
Near each node the coefficients $h_j$ read in low-energy approximation \cite{mccann06}
\beq
h_j=i\nabla_j \ \ (MLG), \ \ h_1=\nabla_1^2-\nabla_2^2, \ h_2=2\nabla_1\nabla_2 \ \ (BLG) \ ,
\label{elements}
\eeq
where $(\nabla_1,\nabla_2)$ is the 2D gradient.

\subsection{Random fluctuations}
\label{randomfluct}

In a realistic situation the potential $V_r$ is not uniform, neither in MLG nor in BLG, as
discussed in the Introduction.
As a result, electrons experience a randomly varying potential $V_r$ along each graphene
sheet, and $m$ in the Hamiltonian of Eq. (\ref{ham01}) becomes a random variable in space as well.
For BLG it is assumed that the gate voltage is adjusted at 
the charge-neutrality point such that in average $m_r$ is 
exactly antisymmetric with respect to the two layers:
$\langle m_1\rangle_m=-\langle m_2\rangle_m$.

At first glance, the Hamiltonian
in Eq. (\ref{ham00}) is a standard hopping Hamiltonian with random potential $V_r$. This is a model
frequently used to study the generic case of Anderson localization \cite{anderson58}. The dispersion,
however, is special in the case of graphene due to the honeycomb lattice: at low energies it 
consists of two nodes (or valleys) $K$ and $K'$ \cite{castroneto07b,mccann06}. 
It is assumed here that randomness scatters only at small momentum such that 
intervalley scattering, which requires large momentum at least near the nodal points (NP), is not relevant
and can be treated as a perturbation.
Then each valley contributes separately to the DOS, and the contribution of 
the two valleys to the DOS $\rho$ is additive:
$
\rho=\rho_K+\rho_{K'}
$.
This allows us to consider the low-energy Hamiltonian in Eqs. (\ref{ham01}), (\ref{elements}),  
even in the presence of randomness for each valley separately. 
Within this approximation the term $m_r$ is a random variable with mean
value $\langle m_r\rangle_m ={\bar m}$ and variance 
$\langle (m_r-{\bar m})(m_{r'}-{\bar m})\rangle_m=g\delta_{r,r'}$.
The following analytic calculations will be based entirely on the Hamiltonian of 
Eqs. (\ref{ham01}),(\ref{elements}) and the numerical calculations on the lattice Hamiltonian
of Eq. (\ref{ham00}). In particular, the average Hamiltonian $\langle H\rangle_m$
can be diagonalized by Fourier transformation and is
\beq
\langle H\rangle_m =
k_1\sigma_1+k_2\sigma_2+{\bar m}\sigma_3
\eeq
 for MLG with eigenvalues $E_k=\pm\sqrt{{\bar m}^2+k^2}$.
For BGL the average Hamiltonian is
\beq
\langle H\rangle_m =
(k_1^2-k_2^2)\sigma_1+2k_1k_2\sigma_2+{\bar m}\sigma_3
\eeq
with eigenvalues $E_k=\pm\sqrt{{\bar m}^2+k^4}$.


\subsection{Symmetries}
\label{symmetry000}

Low-energy properties 
are controlled by the symmetry of the Hamiltonian and of the corresponding 
one-particle Green's function $G(i\epsilon)=(H+i\epsilon)^{-1}$. In the absence of
sublattice-symmetry breaking (i.e. for $m=0$), the Hamiltonian 
$H=h_1\sigma_1+h_2\sigma_2$ has a continuous chiral symmetry
\beq
H \to e^{\alpha\sigma_3} He^{\alpha\sigma_3}=H
\label{contsymmetry}
\eeq
with a continuous parameter $\alpha$, since $H$ anticommutes with $\sigma_3$.
The term $m\sigma_3$ breaks the continuous chiral symmetry. 
However, the behavior under transposition $h_j^T=-h_j$ for MLG and $h_j^T=h_j$ for 
BLG in Eq. (\ref{elements}) provides a discrete symmetry:
\beq
H\to -\sigma_n H^T\sigma_n =H \ ,
\label{discretesymm}
\eeq
where $n=1$ for MLG and $n=2$ for BLG.
This symmetry is broken for the one-particle Green's function $G(i\epsilon)$
by the $i\epsilon$ term. To see whether or not the symmetry is restored in the limit $\epsilon\to0$,
the difference of $G(i\epsilon)$ and the transformed Green's function $-\sigma_nG^T(i\epsilon)\sigma_n$
must be evaluated:
\beq
G(i\epsilon)+\sigma_nG^T(i\epsilon)\sigma_n=G(i\epsilon)-G(-i\epsilon)  \ .
\label{op}
\eeq
For the diagonal elements this is the DOS at the NP $\rho(E=0)\equiv\rho_0$ 
in the limit $\epsilon\to0$.
Thus the order parameter for spontaneous symmetry breaking is $\rho_0$. 
According to the theory of phase transitions, the transition from a nonzero 
$\rho_0$ (spontaneously broken symmetry) to $\rho_0=0$ (symmetric phase)
is a second-order phase transition, and should be accompanied
by a divergent correlation length at the transition point. Since our symmetry is 
discrete, such a phase transition can exists in $d=2$ and should be of Ising type.
A calculation, using the SCBA of $\rho_0$, gives indeed a second-order transition at the
point where $\rho_0$ vanishes with a divergent correlation length $\xi$  for the DOS fluctuations
\[
\xi\sim \xi_0 (m_c^2-{\bar m}^2)^{-1}
\]
for ${\bar m}^2\sim m_c^2$ with a finite coefficient $\xi_0$ \cite{ziegler97}.
Whether or not this transition is an artefact of the SCBA or represents a physical
effect due to the appearence of two types of spectra (localized for vanishing
SCBA-DOS and delocalized for nonzero SCBA-DOS) is not obvious here and requires
further studies.

\subsection{Density of states}

\begin{figure}
\centering
\psfrag{x}[t][b][1][0]{$E/m$}
\psfrag{y}[b][t][1][0]{$m\rho(E)$}
\psfrag{bi}[][][1][0]{\color{red} BLG}
\psfrag{mono}[][][1][0]{\color{blue} MLG}
\includegraphics[width=7cm,height=7cm]{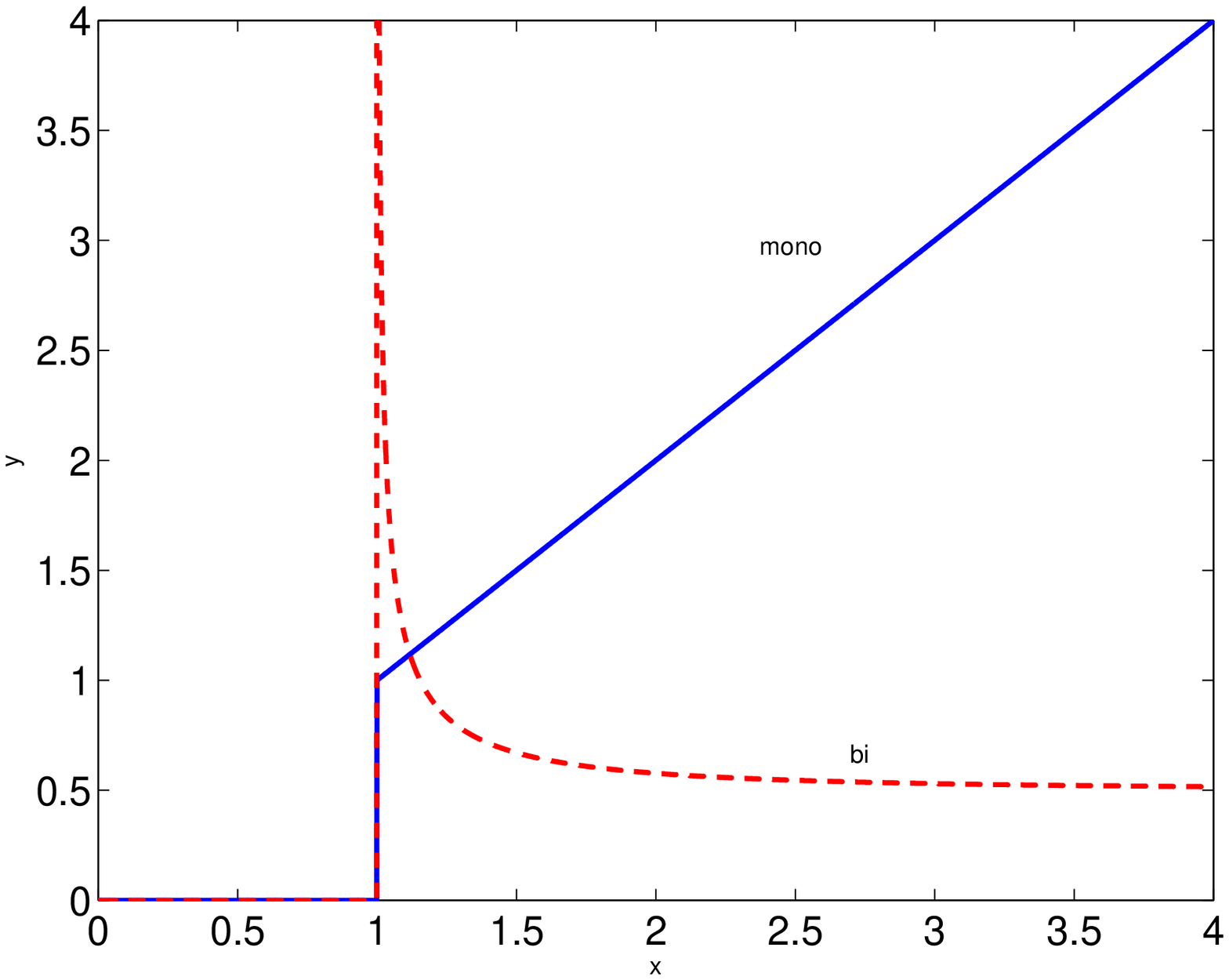}
\hspace*{5mm}
\psfrag{bi1}[][][1][0]{\color{blue} $t_\perp=\infty$}
\psfrag{bi2}[][][1][0]{\color{red}  $t_\perp=2m$}
\psfrag{bi3}[][][1][0]{\color{black}  $2t_\perp=m$}
\psfrag{xx}[t][b][1][0]{$E/m$}
\psfrag{yy}[b][t][1][0]{$t_\perp\rho(E)$}
\includegraphics[width=7cm,height=7cm]{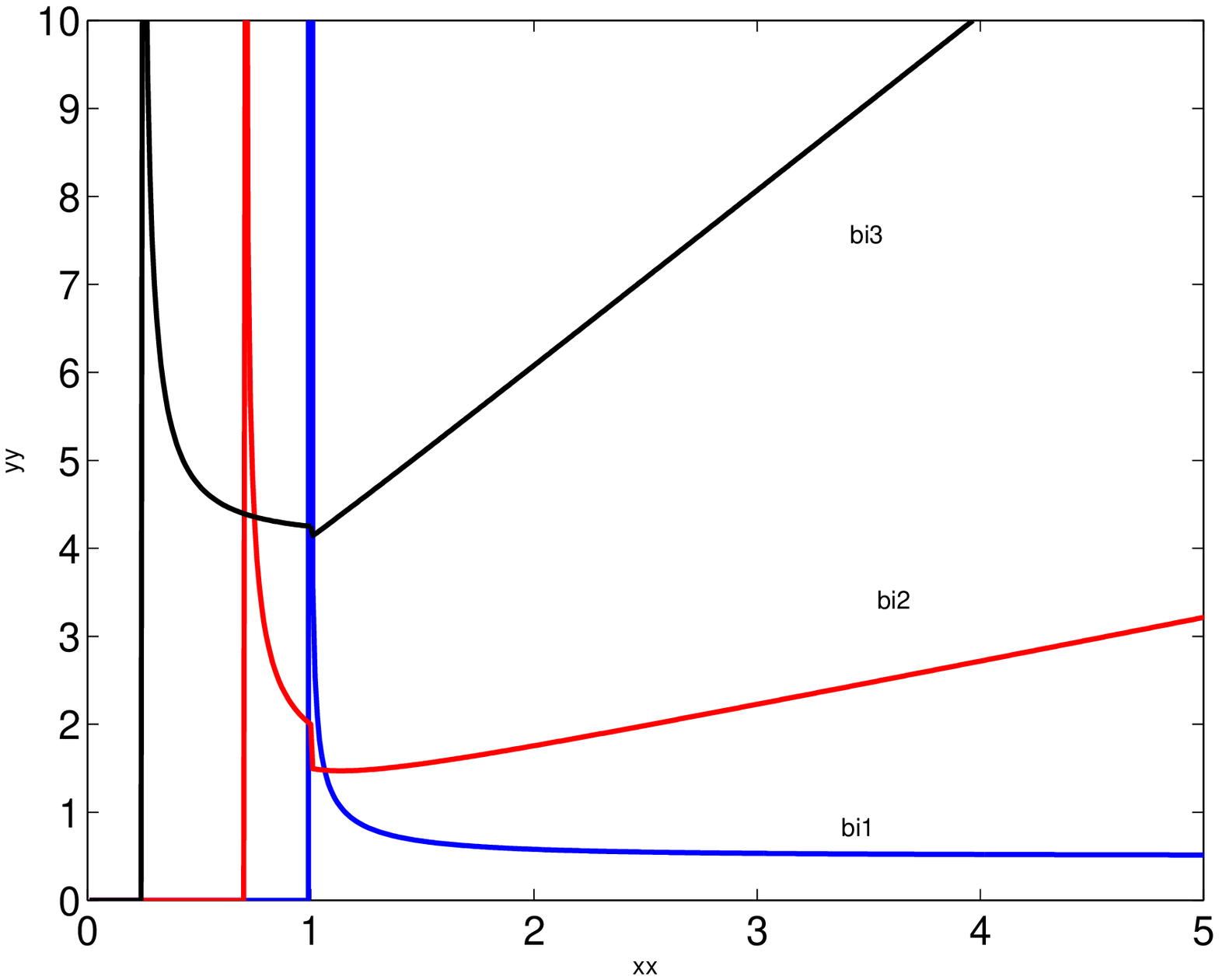}
\caption{Density of states for a uniform symmetry-breaking potential for monolayer
graphene and bilayer graphene is shown in the left panel. 
The density of states for a uniform symmetry-breaking potential for BLG is shown for several values of $t_\perp$. 
For small $t_\perp$, the mexican hat structure influences the DOS by shifting the gap to 
lower values, and by developing a kink at $E=m$.}
\label{dosplot}
\end{figure}

Our focus in the subsequent calculation is on the DOS of MLG and BLG.
In the absence of disorder, the DOS of 2D Dirac fermions opens a gap $\Delta\propto {\bar m}$ 
as soon as a nonzero term ${\bar m}$ appears in the Hamiltonian of Eq. (\ref{ham01}),
since the low-energy dispersion is $E_k=\pm\sqrt{{\bar m}^2+k^2}$ for MLG and
$E_k=\pm\sqrt{{\bar m}^2+k^4}$ for BLG, respectively (cf Fig. \ref{dosplot}).
Here we evaluate the DOS of MLG and BLG in the presence of a uniform gap.
Given the energy spectrum, the DOS is defined as
\begin{equation}
\rho(E)=\sum_k\delta(E-E_k).
\end{equation}
By using the MLG dispersion, this reduces to
\begin{equation}
\rho(E)=|E|\Theta(|E|-m),
\end{equation}
where $\Theta(x)$ is the Heaviside function.
For BLG,
this gives
\begin{equation}
\rho(E)=\frac{|E|}{2\sqrt{E^2-m^2}}\Theta(|E|-m),
\label{dosbil}
\end{equation}
which are shown in Fig. \ref{dosplot}.
By retaining the full low-energy spectrum for BLG, $E_k^-$, the DOS can still be evaluated in closed form, with the result
\begin{eqnarray}
\fl \rho(E)=|E|\times \left\{\begin{array}{cc}
\frac{(t_\perp^2+4m^2)}{\sqrt{(t_\perp^2+4m^2)E^2-t_\perp^2m^2}}  & \textmd{for } m>|E|>\frac{mt_\perp}{\sqrt{t_\perp^2+4m^2}}\\
\left(\frac{(t_\perp^2+4m^2)}{2\sqrt{(t_\perp^2+4m^2)E^2-t_\perp^2m^2}}+1\right)  & \textmd{for }|E|>m.\\
\end{array}\right.
\end{eqnarray}
In the limit of $t_\perp\gg (E,m)$, this reduces to Eq. (\ref{dosbil}) after dividing it by $t_\perp$, which was set to 1 in the
low-energy approximation, and the DOS
saturates to a constant value after the initial divergence. For finite $t_\perp$, however, the Dirac nature of the spectrum appears again, 
and the high energy DOS increases linearly even for the BLG, 
similarly to the MLG case.
For $m=0$, and $E\ll t_\perp$, this lengthy expression gives
\begin{equation}
\rho(E\ll t_\perp)=\frac{t_\perp}{2}.
\end{equation}

An interesting question, from the theoretical as well as from the experimental point of view,
appears here:
What is the effect of random fluctuations around ${\bar m}$? Previous calculations,
based on the self-consistent Born approximation (SCBA), have revealed that those fluctuations can close
the gap again, even for an average SBP term ${\bar m}\ne0$ \cite{ziegler09b}. Only if ${\bar m}$ exceeds
a critical value $m_c$ (which depends on the strength of the fluctuations), an open gap was found in these
calculations. This describes a special transition from metallic to insulating behavior.
In particular, the DOS at the Dirac point $\rho_0$ vanishes with ${\bar m}$ like a power law
\beq
\rho_0({\bar m})\sim \sqrt{{\bar m}-m_c^2} \ .
\eeq
The exponent 1/2 of the power law is probably an artefact of the SCBA, similar to the critical exponent in
mean-field approximations.

\section{Self-consistent Born approximation}

The average one-particle Green's function can be calculated from the average Hamiltonian 
$\langle H\rangle_m$ by employing the self-consistent Born approximation (SCBA)
\cite{suzuura02,peres06,koshino06}
\beq
\langle G(i\epsilon)\rangle_m\approx (\langle H\rangle_m+i\epsilon - 2\Sigma)^{-1}
\equiv G_0(i\eta,m_s) \ .
\label{scba1}
\eeq
The SCBA is also known as the self-consistent non-crossing approximation in the Kondo and superconducting community.
The self-energy $\Sigma$ is a $2\times2$ tensor due to the spinor structure
of the quasiparticles: $\Sigma=-(i\eta\sigma_0+m_s \sigma_3)/2$.
Scattering by the random SBP 
produces an imaginary part of the self-energy $\eta$ (i.e. a one-particle scattering rate)
and a shift $m_s$ of the average SBP ${\bar m}$ (i.e., ${\bar m}\to m'\equiv {\bar m}+m_s$). 
$\Sigma$ is determined by the self-consistent equation
\beq
\Sigma=-g\sigma_3(\langle H\rangle_m+i\epsilon  -2\Sigma)^{-1}_{rr}\sigma_3 \ .
\label{spe00}
\eeq
The symmetry in Eq. (\ref{discretesymm}) implies that with $\Sigma$ also 
\beq
\sigma_n\Sigma\sigma_n=-(i\eta\sigma_0-m_s \sigma_3)/2
\eeq
is a solution (i.e. $m_s\to -m_s$ creates a second solution).

The average DOS at the NP is proportional to the scattering rate:
$\rho_0=\eta/2g\pi$. This reflects that scattering by the random SBP 
creates a nonzero DOS at the NP if $\eta>0$. 

Now we assume that the parameters $\eta$  and $m_s$ are uniform in space. 
Then Eq. (\ref{spe00}) can be written in 
terms of two equations, one for the one-particle scattering rate $\eta$ and 
another for the shift of the SBP $m_s$, as 
\beq
\eta= gI\eta, \ \ m_s=-{\bar m} gI/(1+gI) \ .
\label{scba2}
\eeq
$I$ is a function of ${\bar m}$ and $\eta$ and also depends on the Hamiltonian. 
For MLG it reads with momentum cutoff $\lambda$
\begin{equation}
I_{MLG}= 
\frac{1}{2\pi}\ln\left[ 1+\frac{\lambda^2}{{\eta}^2 +({\bar m}+m_s)^2}\right]
\label{int1}
\end{equation}
and for BLG
\begin{equation}
I_{BLG}\sim \frac{1}{4\sqrt{{\eta}^2+({\bar m}+m_s)^2}}\ \ \ \ (\lambda\sim\infty) \ .
\label{int2}
\end{equation}
A nonzero solution $\eta$ requires $gI=1$ in the first part of Eq. (\ref{scba2}), 
such that $m_s=-{\bar m}/2$ from the second part. 
Since the integrals $I$ are monotonically decreasing functions for large ${\bar m}$,
a real solution with $gI=1$ exists only for $|{\bar m}|\le m_c$. For both, MLG and BLG,
the solutions read
\beq
\eta^2=(m_c^2-{\bar m}^2)\Theta(m_c^2-{\bar m}^2)/4 \ ,
\label{scattrate}
\eeq
where the model dependence enters only through the critical average SBP $m_c$:
\beq
m_c=\left\{\begin{array}{cc}
(2\lambda /\sqrt{e^{2\pi/g}-1})\sim 2\lambda e^{-\pi/g} & MLG \\
g/2 & BLG 
\end{array}\right.
.
\label{gap11}
\eeq
$m_c$ is much bigger for BGL, a result which indicates that the effect of disorder 
is much stronger in BLG. This is also reflected by the scattering rate at ${\bar m}=0$ which 
is $\eta=m_c/2$.

A central assumption of the SCBA is a uniform self-energy $\Sigma$. The imaginary part
of $\Sigma$ is the scattering rate $\eta$, created by the random fluctuations. Therefore, 
a uniform $\eta$ means that effectively random fluctuations are densely filling the lattice. 
If the distribution of the fluctuations is too dilute, however,
there is no uniform nonzero solution of Eq. (\ref{spe00}). Nevertheless, a dilute distribution
can still create a nonzero DOS, as we will discuss in the following: we study contributions 
to the DOS due to rare events, leading to Lifshitz tails. 

\begin{figure}
\begin{center}
\includegraphics[width=7cm,height=5cm]{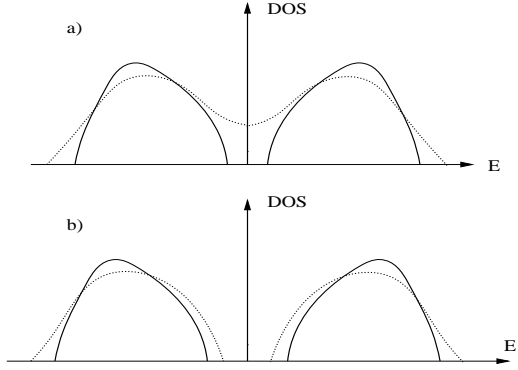}
\caption{
Schematic shape of the density of states: full curves are the bulk density of states for uniform 
symmetry-breaking potential, dotted
curves represent the broadening by disorder. The broadened density of states can overlap inside the 
gap for ${\bar m}<m_c$ (a) or not for ${\bar m}>m_c$ (b), 
depending on the average symmetry-breaking potential ${\bar m}$. $m_c$ is given in Eq. (\ref{gap11}).
}
\label{dosplot1}
\end{center}
\end{figure}

\section{Lifshitz tails}
\label{lifshitztails}

In the system with uniform SBP the gap can be destroyed locally by a local change of the SBP
$m\to m+\delta m_r$ due to the creation of a bound state. 
We start with a translational-invariant system and add $\delta m_r$ on site $r$.
To evaluate the corresponding DOS from the Green's function $G=(H+i\epsilon +\delta m\sigma_3)^{-1}$, 
using the Green's function $G_0=(H+i\epsilon)^{-1}$ with uniform $m$,
we employ the lattice version of the Lippmann-Schwinger equation \cite{ziegler85}
\beq
G=G_0-G_0T_SG_0=({\bf 1}-G_0T_{r})G_0 
\label{id3}
\eeq
with the $2\times2$ scattering matrix
\beq
T_{r}=(\sigma_0+\delta m_{r}\sigma_3 G_{0,rr})^{-1}\sigma_3\delta m_{r} \ .
\label{impurity00}
\eeq
In the case of MLG we have
\begin{eqnarray}
G_0=\left[(E+i\epsilon)\sigma_0-m\sigma_3\right]\frac{1}{2\pi}
\int_0^\lambda \frac{k}{(\epsilon-iE)^2+m^2+k^2}dk\\
\sim (E\sigma_0-m\sigma_3)\frac{1}{4\pi}\log[1+\lambda^2/(m^2-E^2)]+o(i\epsilon)
\equiv (g_0+i\epsilon s)\sigma_0+g_3\sigma_3 \ .
\end{eqnarray}
(remark: the DOS of BLG has the same form.) Then the imaginary part of the 
Green's function reads
\begin{eqnarray}
Im[G(\eta)]=-\left(\begin{array}{cc}
\delta_{\epsilon s}(g_0+g_3+\delta m_{r}) & 0 \\
0 & \delta_{\epsilon s}(g_0-g_3-\delta m_{r}) 
\end{array}\right)
\end{eqnarray}
with
\beq
\delta_{\epsilon s}(x)=\frac{\epsilon s}{x^2+\epsilon^2 s^2} \ .
\eeq
Thus the DOS is the sum of two Dirac delta peaks
\beq
\rho_r\propto \delta_{\epsilon s}(g_0+g_3+\delta m_{r})+ \delta_{\epsilon s}(g_0-g_3-\delta m_{r}) \ .
\eeq
The Dirac delta peak appears with probability $\propto\exp(-(g_0\pm g_3)^2/g)$ for a Gaussian
distribution. This calculation can easily be generalized to $\delta m_r$ on a set of several 
sites $r$ \cite{ziegler85}. Then the appearance of the several such Dirac delta peaks decreases 
exponentially. Moreover, these contributions are local and form localized states.  
For stronger fluctuations $\delta m_r$ (i.e., for increasing $g$) the localized states
can start to overlap. This is a quantum analogue of classical percolation.

The localized states in the Lifshitz tails can be taken into account by a generalization of the SCBA
to non-uniform self-energies. The main idea is to search for space-dependent solutions 
$\Sigma_r$ of Eq. (\ref{spe00}). In general, this is a diffult problem. However, we have found that
this problem simplifies essentially when we study it in terms of a $1/{\bar m}$ expansion.
Using a Gaussian distribution, this method gives Lifshitz tails of the form \cite{villain00}:
\beq
\rho_0({\bar m}) 
\sim  \frac{{\bar m}^4}{ 32\sqrt{\pi}g^{5/2}}e^{-{\bar m}^2/4g} \ .
\eeq

\section{Numerical approach}

To understand to behavior of random gap fluctuations in graphene, and also the limitations of the SCBA,
we carried out extensive numerical simulations on the honeycomb lattice, allowing for various random gap fluctuations on top of a 
uniform gap $m$. These fluctuations are simulated by box and Gaussian distributions.
From the SCBA, the emergence of a second-order phase transition at a critical mean $m_c$ is obvious for a given 
variance. This is best manifested in the behavior of the DOS, which stays finite for $\langle m\rangle <m_c$, and 
vanishes afterwards, and serves as an order parameter.
Does this picture indeed survive, when higher order corrections in the fluctuations are taken into account?

\begin{figure}[h!]
\centering
\psfrag{x}[t][b][1][0]{$m/t$}
\psfrag{y}[b][t][1][0]{eigenvalues$/t$}
\psfrag{g1}[][][1][0]{$g=0.6^2$}  
\psfrag{g2}[][][1][0]{$g=0.8^2$}
\psfrag{g3}[][][1][0]{$g=1$}
\includegraphics[width=7cm,height=10cm]{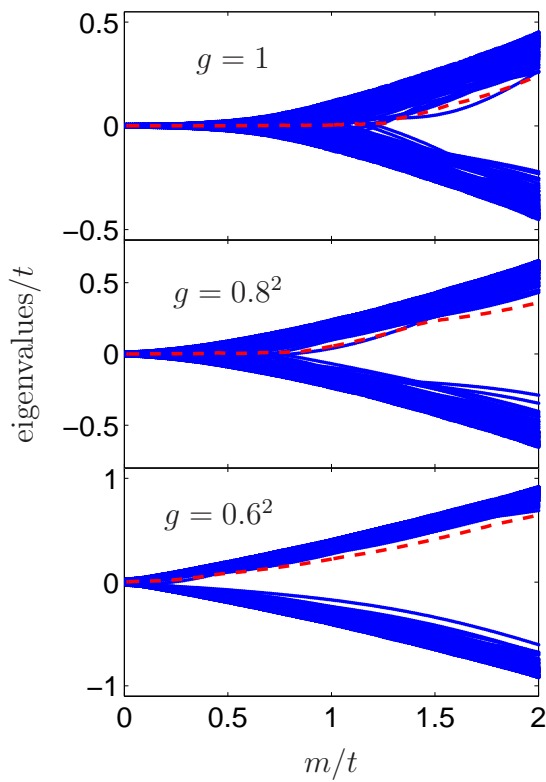}
\caption{(Color online) The evolution of the 200 lowest eigenvalues is shown for a given random mass configuration with Gaussian
distribution (with variance g) on a 600x600 HCL, by varying the uniform gap. The red line denotes the maximum of the level 
spacing of these eigenvalues, a possible definition of the average gap.
\label{rmeigenval600}}
\end{figure}

To start with, we take a fix random mass  configuration with a given variance and the honeycomb lattice (HCL) with the conventional hoppings ($t$), represented 
by $H_0$. Then, we take a separate Hamiltonian, responsible for the uniform, non-fluctuating gaps, denoted by $H_{gap}$, and study the 
evolution   of the eigenvalues of $H_0+mH_{gap}$ by varying m for a 600x600 lattice. 
By using Lanczos diagonalization, we focus our attention only to the 200 eigenvalues closest to the NP. 
Their evolution is shown in Fig. \ref{rmeigenval600}.
This supports the existence of a finite $m_c$, but since it originates from a single random disorder configuration, rare events can 
alter the result. As a possible definition of the rigid gap, we also show the maximum of the energy level spacing for these eigenvalues as a function of $m$.
As seen, it starts to increase abruptly at a given value of $m$, which can define $m_c$.

\begin{figure}[h!]
\centering
\psfrag{x}[t][b][1][0]{$m/t$}
\psfrag{y}[b][t][1][0]{$\rho(0)t$}
\psfrag{xx}[t][b][1][0]{$g$}
\psfrag{yy}[b][t][1][0]{exponent ($c$)}
\includegraphics[width=8cm,height=8cm]{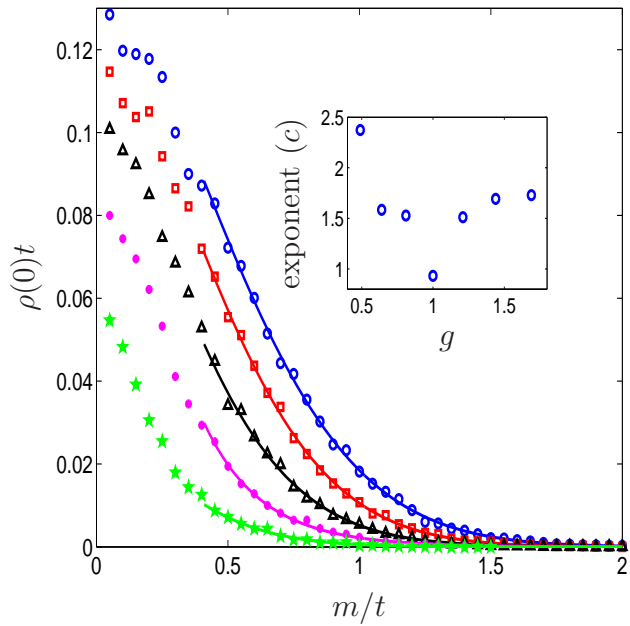}
\caption{(Color online) The density of states at the NP is plotted for Gaussian random mass for a 200x200 HCL for $g=0.9^2$, 1,
$1.1^2$, $1.2^2$ and $1.3^2$ from bottom to top after 400 averages. The symbols denote the numerical data, solid lines are fits using
$a\exp(-b m^c)$. The inset shown the obtained exponents, $c$, as a function of $g$, which is close to 1.5.
\label{rmgauss}}
\end{figure}

To investigate whether a finite critical $m_c$ survives,
 we take smaller systems and evaluate the averaged DOS directly  from many disorder realizations. To achieve this, we take a 200x200 
HCL, and evaluate the 200 closest eigenvalues to the NP, and count their number in a given small interval, $\Delta E$ 
(smaller 
than the maximal eigenvalues) around zero. This method was found to be efficient in studying other types of randomness \cite{ziegler08b}.
We mention that large values of $\Delta E$ take contribution from higher energy states into account, while too small values are 
sensitive to the discrete lattice and consequently the discrete eigenvalue structure of the Hamiltonian. For lattices containing a 
few $10^4-10^5$ sites, $\Delta E/t\sim 10^{-2}-10^{-4}$ are convenient.

\begin{figure}[h!]
\centering
\psfrag{x}[t][b][1][0]{$m/t$}
\psfrag{y}[b][t][1][0]{$\rho(0)t$}
\psfrag{xx}[t][b][1][0]{$g$}
\psfrag{yy}[b][t][1][0]{exponent ($c$)}
\includegraphics[width=8cm,height=8cm]{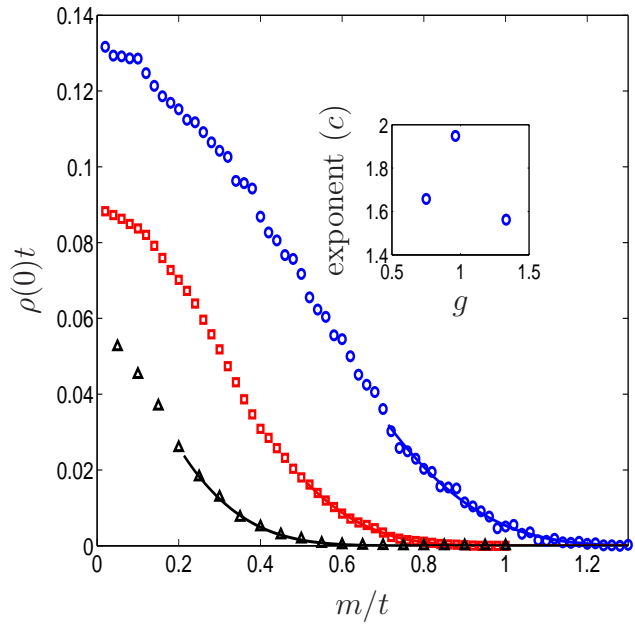}
\caption{(Color online) The density of states at the NP is plotted for box distributed ($[-W..W]$) randomness
for a 200x200 HCL for $W=1.5$, 1.7 and 2 ($g=W^2/3$) from bottom to top after 400 averages. 
The symbols denote the numerical data, solid lines are fits using
$a\exp(-b/|m-W|^c)$. The inset shown the obtained exponents, $c$, as a function of $g$.
\label{rmbox}}
\end{figure}

The resulting DOS is plotted in Figs. \ref{rmgauss} and \ref{rmbox} for Gaussian (with variance $g$) and box distribution (within $[-W..W]$, variance $g=W^2/3$). This does not indicate a sharp threshold, 
but rather the development of long Lifshitz tails due to randomness, as we already predicted in the previous section.
To analyze them, we fitted the numerical data by assuming exponential tails of the form
\begin{equation}
\rho(0)=t\exp\left(-a-b m^c\right)
\end{equation}
for a Gaussian and
\begin{equation}
\rho(0)=t\exp\left(-a-b/|m-W|^c\right) 
\end{equation}
for a box distribution, as suggested by Ref. \cite{Cardy}.
The obtained $c$ values are  visualized in the insets of Figs. \ref{rmgauss} and \ref{rmbox}.
Given the good agreement, we believe that the DOS at the NP is made of states that are localized in a Lifshitz tail.
We mention that these results are not sensitive to finite size scaling at these values of the disorder and uniform gap, 
only smaller systems (like the 30x30 HCL) require more averages ($\sim 10^4$), whereas for larger ones (such as the 200x200 with 
400 averages) fewer averages are sufficient.

\begin{figure}[h!]  
\centering
\psfrag{x}[t][b][1][0]{$E/t$}
\psfrag{y}[b][t][1][0]{$\rho(E)t$}
\psfrag{xx}[t][b][1][0]{$(E/t)^2\sim n$}
\psfrag{yy}[b][t][1][0]{$1/\rho(E)t\sim T_0$}
\includegraphics[width=7cm,height=7cm]{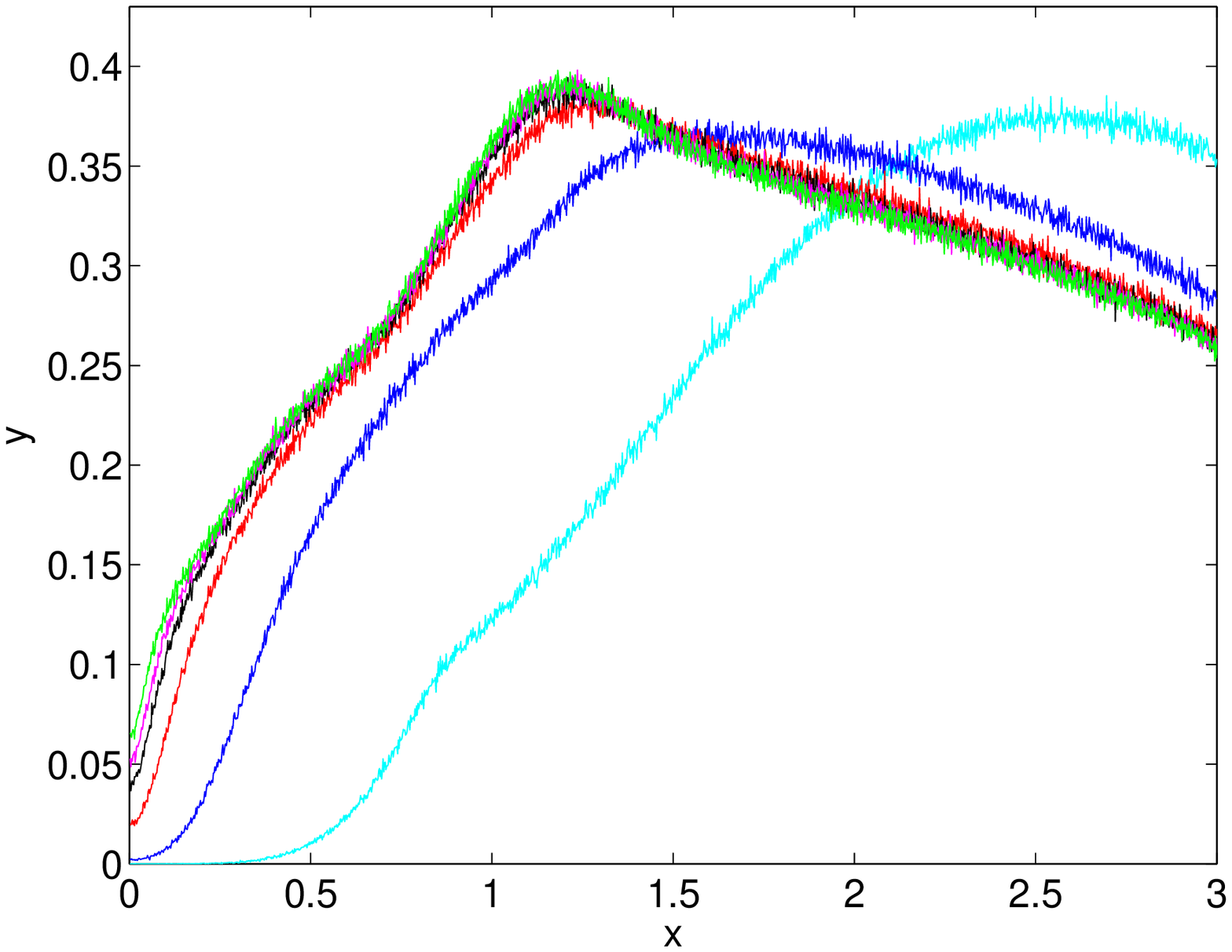}
\hspace*{4mm}
\includegraphics[width=7cm,height=7cm]{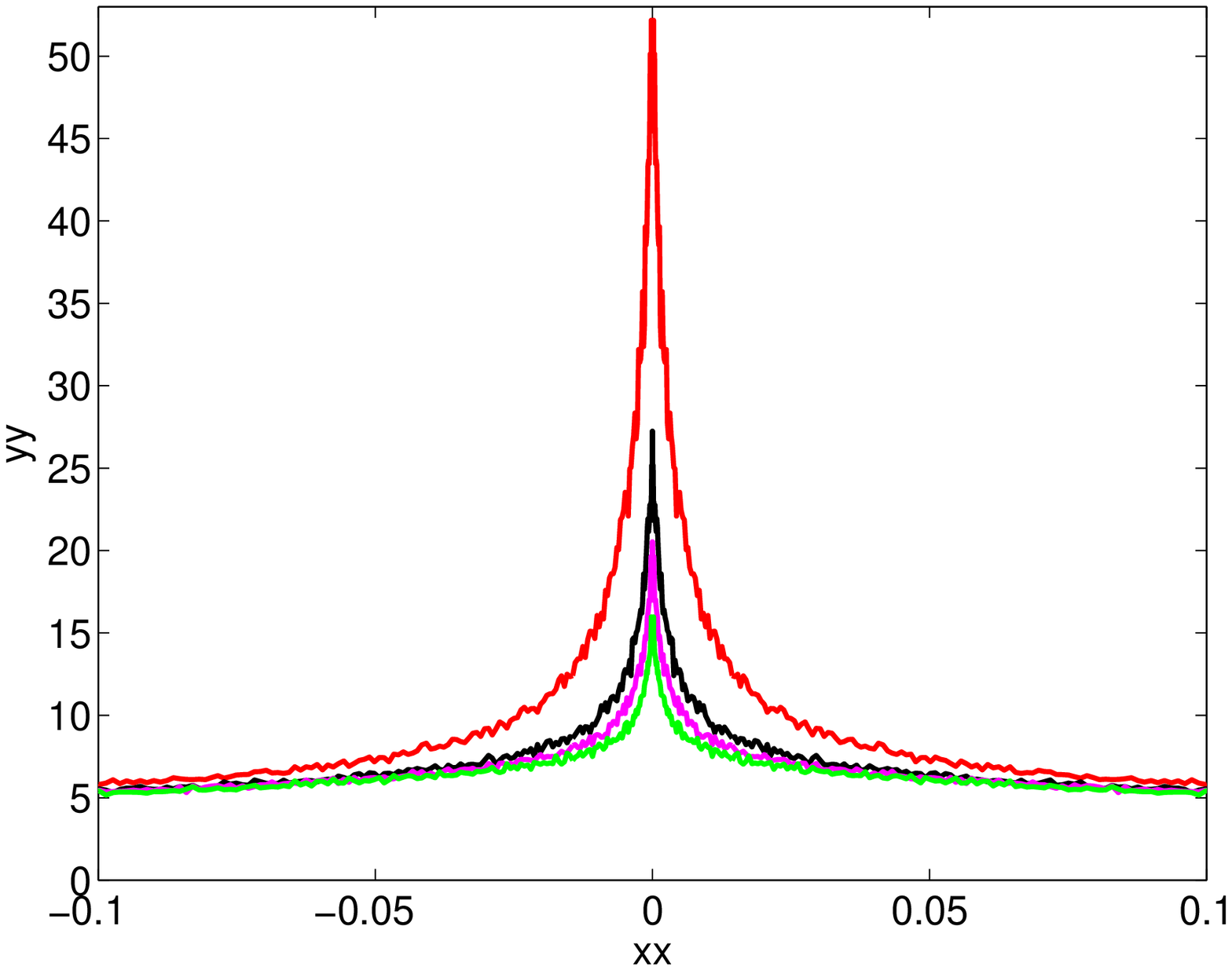}
\caption{(Color online) The energy dependent density of states is plotted for Gaussian distributed random mass for a 30x30 HCL after 
$10^4$ averages for $g=1$, $m=2$ (cyan), 1 (blue), 0.5 (red), 0.3 
(black), 0.2 (magenta) and 0 (green) in the left panel.
The right panel visualizes the inverse of the density of states, being proportional to $T_0$ in the variable range hopping model as a 
function of the energy squared (proportional to the carrier density). 
\label{dosenerg}}
\end{figure}

In Fig. \ref{dosenerg}, the energy dependent DOS is shown for Gaussian random mass with $g=1$ and for several uniform gap values. 
With increasing $m$, the DOS dimishes rapidly at low energies, and develops a pseudogap. The logarithmic singularity at $E=t$ is washed out for $g=1$.
We also show the inverse of the DOS, proportional to $T_0$, the characteristic temperature scale of variable range hopping as 
a function of the carrier density (which is proportional to $E^2$).

\section{Discussion}

MLG and BLG consist of two bands that touch each other at two nodal points (or valleys).
Near the nodes the spectrum of MLG is linear (Dirac-like) and the spectrum of BLG is quadratic.
The application of a uniform SBP opens a gap in the DOS for both cases. For a random SPB, 
however, the situation is less obvious. First of all, it is clear that randomness
leads to a broadening of the bands. If we have two separate bands due to
a small uniform SPB, randomness can close the gap again due to broadening (cf. Fig. \ref{dosplot}a).
The broadening of the bands depends on the strength of the fluctuations of the random SBP. 
In the case of a Gaussian distribution there are energy tails for all energies. 

Now we focus on the NP, i.e. we consider $E=0$ and $\rho_0$. 
Then we have two parameters in order to change the gap structure: the average SBP
$\langle m\rangle\equiv{\bar m}$ and the variance $g$. ${\bar m}$ allows us to broaden the
gap and $g$ has the effect of closing it due to broadening of the two subbands.
Previous calculations have shown that at the critical value $m_c(g)$ of Eq. (\ref{gap11})
the metallic behavior breaks down for ${\bar m}>m_c(g)$ \cite{ziegler09b}. On the other hand,
Gaussian randomness creates tails at all energies. Consequently, there are localized states
for $|{\bar m}|\ge m_c(g)$ at the NP, and there are delocalized states for 
$|{\bar m}|<m_c(g)$ at the NP.
The localized states in the tails are described, for instance, by the Lippmann-Schwinger 
equation (\ref{id3}) . The SCBA with uniform self-energy is not able to produce the
localized tails. An extension of the SCBA with non-uniform self-energies provides localized
tails though, as an approximation for large ${\bar m}$ has shown \cite{villain00}.
This is also in good agreement with our exact diagonalization of finite systems up to 
$200\times200$ size.

A possible interpretation of these results is that there are two different types of spectra.
In a special realization of $m_r$ the tails of the DOS represent localized states. On the
other hand, the DOS at the NP $E=0$, obtained from the SCBA with {\it uniform}
self-energy, comes from extended states \cite{ziegler09b}. The localized and the delocalized 
spectrum separate at the critical value $m_c$, undergoing an Anderson transition.

{\it Conductivity}:
Transport, i.e. the metallic regime, is related to the DOS trough the Einstein relation 
$\sigma\propto \rho D$, where $D$ is the diffusion coefficient. The latter was found in Ref. \cite{ziegler09b}
for $E\sim0$ as
\beq
D=\frac{ag\sqrt{m_c^2-{\bar m}^2}}{2\pi m_c^2}\Theta(m_c^2-{\bar m}^2) \ , 
\label{diffcoeff3}
\eeq
where $a=1$ ($a=2$) for MLG (BLG). Together with the DOS $\rho_0=\eta/2g\pi$ and the scattering rate $\eta$
in Eq. (\ref{scattrate}), the Einstein relation gives us at the NP
\begin{equation}
\sigma(\omega\sim0)\propto \rho_0 D\frac{e^2}{h}
\approx \frac{a}{8\pi^2}\left(1-\frac{{\bar m}^2}{m_c^2}\right)\Theta(m_c^2-{\bar m}^2)\frac{e^2}{h}  \ .
\end{equation}

In the localized regime (i.e. for $|{\bar m}|\ge m_c$) the conductivity is nonzero only for
positive temperatures $T>0$. Then we can apply the formula for variable-range hopping in Eq. (\ref{vrh}),
which fits well the experimental result in graphane of Ref. \cite{elias08}.
The parameter $T_0$ is related to the DOS at the Fermi level as \cite{mott69}
\begin{equation}
k_B T_0\propto\frac{1}{\xi^2 \rho(E_F)} \  ,
\end{equation}
where $\xi$ is the localization length. 
$T_0$ has its maximum at the NP $E_F=0$, as shown in Fig. \ref{dosenerg} and decreases
monotonically with increasing carrier density, as in the experiment on graphane
\cite{elias08}.

In conclusion, we have studied the density of states in MLG and BLG at low energies in the presence of
a random symmetry-breaking
potential. While a uniform symmetry-breaking potential opens a uniform gap, a random symmetry-breaking potential
also creates tails in the density of states. The latter can close the gap again, preventing the system to become an
insulator at the nodes. However, for a sufficiently large gap the tails contain localized states with nonzero density
of states. These localized states allow the system to conduct at nonzero temperature via variable-range hopping.
This result is in agreement with recent experimental observations \cite{elias08}.    

\vskip0.5cm
\noindent
{\bf Acknowledgement:}
This work was supported by a grant from the Deutsche Forschungsgemeinschaft and  by the Hungarian
Scientific Research Fund under grant number K72613.

\end{document}